\documentclass[preprint,12pt]{elsarticle}

\usepackage{amssymb}
\usepackage{amsmath}
\usepackage{graphicx}
\usepackage{dcolumn}
\usepackage{bm}
\usepackage{braket}

\newcommand{\bea}{\begin{eqnarray}}
\newcommand{\eea}{\end{eqnarray}}
\newcommand{\simgt}{\hbox{ \raise3pt\hbox to 0pt{$>$}\raise-3pt\hbox{$\sim$} }}
\newcommand{\simlt}{\hbox{ \raise3pt\hbox to 0pt{$<$}\raise-3pt\hbox{$\sim$} }}

\newcommand{\LQ}{\Lambda_{\rm QCD}}

\newcommand{\be}{\begin{equation}}
\newcommand{\ee}{\end{equation}}

\journal{Physics Letter B}

\begin{document}

\begin{frontmatter}



\title{ 
UV contribution and
power dependence on $\LQ$ \\
of Adler function}


\author{G.~Mishima$^\dagger$, Y.~Sumino$^\ast$ and H.~Takaura$^\ast$}

\address{$^\dagger$ Department of Physics, University of Tokyo
Bunkyo-ku, Tokyo, 113-0033 Japan
\\
$^\ast$Department of Physics, Tohoku University,
Sendai, 980--8578 Japan
}

\begin{abstract}
We formulate a way to separate UV and IR contributions
to the Adler function 
and discuss how $\LQ^2/Q^2$ dependence is encoded in the UV contribution
within perturbative QCD.
\end{abstract}


\end{frontmatter}



Perturbative QCD has made remarkable progress
in recent years.
Thanks to developments in computational technology,
the first few to several terms of perturbative series
have become available for
a number of physical quantities.
Due to severe infrared (IR) divergences inherent in
perturbative QCD,
it has become a standard procedure in many of these computations 
to factorize ultraviolet (UV) and IR 
contributions \cite{Collins:2011zzd}.
As more accurate predictions became available,
it is also becoming practically important to factorize
IR renormalons, in addition to IR divergences.
An IR renormalon
reflects the IR structure of an observable
in terms of perturbative QCD and
induces a diverging behavior of the perturbative
series.
For observables which permit operator product expansion
(OPE), factorization can be carried out
more systematically.
In this case, an IR renormalon in a Wilson coefficient
induces a perturbative uncertainty of the same order
of magnitude as the associated nonperturbative
matrix element.
This makes it necessary to subtract the IR renormalon
from the perturbative evaluation of the
Wilson coefficient and to absorb it into
the matrix element, which also agrees with the concept
of Wilsonian approach.
So far this procedure has not been formulated 
completely, and such a formulation is requisite
for precision analyses of QCD in near future.

The Adler function is defined from the derivative
of the hadronic vacuum polarization of the photon. 
It was originally introduced for a phenomenological analysis of
the muon anomalous magnetic moment and the spectra of muonic atoms \cite{Adler:1974gd}.
Since then, it has been playing an important role in precise calculations of 
the muon anomalous magnetic moment, 
running of the QED coupling constant $\alpha _\mathrm{QED} (k)$ from $k=0$ to $M_Z$ \cite{Hagiwara:2011af},
$R$ ratio in $e^+e^-$ collision \cite{Baikov:2008jh},
the inclusive $\tau$ lepton hadronic decay \cite{Nesterenko:2013vja}, etc.
Furthermore, the Adler function serves as an ideal laboratory for
various theoretical tests.
For instance, dispersion relation, sum rules, 
lattice QCD calculations,
perturbative QCD predictions, renormalons, 
various models of IR physics, 
predictions of supersymmetric QCD, etc.,
have been examined.

According to the analysis of IR renormalons, 
the perturbative series of the Adler function contains a renormalon
which induces an
order $\LQ^4/Q^4$ uncertainty.
That is, it has the same dimension as 
the leading nonperturbative matrix element given
by the local gluon condensate \cite{Mueller:1984vh}.
Recently 
perturbative series of the plaquette on lattice, which is similar to the Adler function
in the continuum limit, has been computed up to 35 loops 
and a renormalon behavior of order $\LQ^4/Q^4$ was observed \cite{Bali:2014fea}. 
It has the same order of magnitude as
the gluon condensate and
the observation supports 
our understanding that IR renormalons appear
with the same dimensions as the nonperturbative matrix elements.
However, this does not mean that we understand all the power
corrections $\sim (\LQ/Q)^n$~\cite{Beneke:1998ui}.
There may be power corrections which originate from UV contributions.
In OPE Wilson coefficients may contain power corrections.
To predict each Wilson coefficient accurately
it is important to subtract IR renormalons from the
perturbative series of the Wilson coefficient.
This concurrently
defines the associated nonperturbative matrix element
accurately.

\begin{figure}[t]
 \begin{center}
 \includegraphics[width=0.9\linewidth]{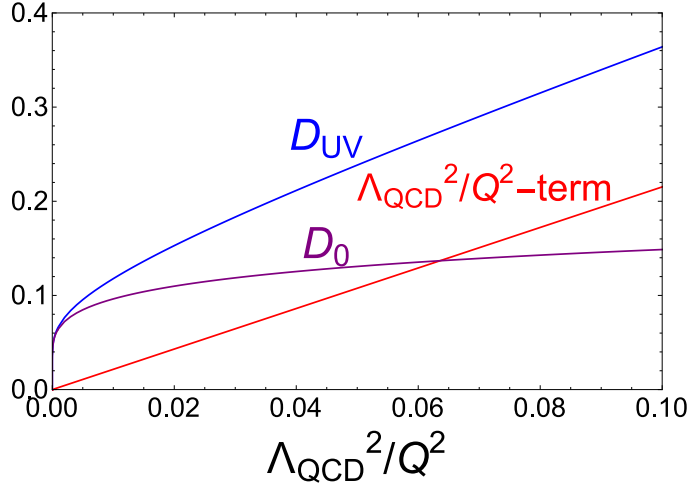}
 \vspace*{-1mm}
 \caption{\small{$D_{\rm UV}$, $D_0$ and the $\LQ^2/Q^2$ term 
 of eq.~(\ref{defDUV}) as functions of $\LQ^2/Q^2$}. }
 \vspace*{-8mm}
 \label{Fig.Linear}
 \end{center}
 \end{figure}

In this Letter we formulate a method to extract UV contributions
to the Adler function, which can be used in OPE.\footnote{
In conventional analyses of renormalons, 
a UV scale is assumed to be much larger than 
any scale involved in the calculation. 
In this Letter, however, we use the terminology ``UV'' for scales
above the factorization scale $\mu_f$ in 
the context of OPE.
In particular $Q$ is regarded as a UV scale.
}
Related subjects have been studied in 
\cite{Mueller:1984vh,Beneke:1994qe,Neubert:1994vb}
(see also~\cite{Narison:2001ix,Chetyrkin:1998yr}), in which
UV and IR contributions have been separated
and their nature has been elucidated.
It was shown that IR contributions induce
order $\LQ^4/Q^4$ renormalon uncertainty to the
perturbative prediction and an explicit
integral representation has been given for the
UV contribution \cite{Ball:1995ni}.
Existence of a $\LQ^2/Q^2$ dependence in the UV contribution
has been discussed, e.g., using a resummation of the perturbative 
series \cite{Ball:1995ni}, and in certain model calculations \cite{Narison:2001ix,Chetyrkin:1998yr}.
Our work can be regarded as an extension of 
the analyses in
refs.~\cite{Beneke:1994qe,Neubert:1994vb,Ball:1995ni}.
We study the (reduced) Adler function $D_{\beta_0}$ with
an explicit IR cut-off $\mu_f$
\cite{Neubert:1994vb}, and in
the large-$\beta_0$ approximation \cite{Beneke:1994qe}.
It gives a natural definition of the Wilson coefficient
of the Adler function based on the Wilsonian picture.
We show that there exists a genuine UV part, which satisfies
$D_{\rm UV}=D_{\beta_0}(\mu_f) + {\cal O}(\mu_f^4/Q^4)$
and is independent of $\mu_f$.
Furthermore, $D_{\rm UV}$ can be expressed as a sum
of a logarithmic term\footnote{
By a ``logarithmic term'' we mean a term 
which is closest
to $(Q^2/\LQ^2)^P$ with $P=0$
in the entire range $0<Q^2<\infty$, if it is
compared with a single power dependence on $Q^2$
(for an integer $P$);
see eq.~(\ref{asympt-Dp}) and Fig.~\ref{Fig.Linear}.
}
and a $\LQ^2/Q^2$ term.
We also discuss its scheme dependence.
We believe that these add information to our
previous knowledge, and moreover,
the formulation provides a simple and
clear picture which would be useful in accurate OPE analyses.

We adapt a formulation used in the analysis of the static QCD potential
\cite{Sumino:2003yp,Sumino:2004ht,Sumino:2014qpa}
after appropriate modifications.
In the case of the static potential a ``Coulomb+linear'' form
(with logarithmic correction at short-distances)
is extracted as the UV contribution,
which reproduces lattice
results at $r\lesssim 0.25$~fm.
This feature provides a guide to our analysis
of the UV contribution to the Adler function,
in particular concerning power dependence on $\LQ/Q$.

We define the reduced Adler function by 
\bea
D(Q^2)= 4 \pi^2 Q^2 \frac{d \Pi(-Q^2)}{d Q^2}-1\,,
~~~ Q^2=-q^2 \,.
\label{eq:defAdlerFn}
\eea
[The Adler function is given by $1+D(Q^2)$ up to a convention
dependent normalization factor.]
$\Pi(q^2)$ denotes the hadronic vacuum polarization given by
\be
i \!\int\! d^4 x \,  e^{i q \cdot x} 
\braket{0|TJ^{\mu}(x) J^{\nu}(0)|0}=(q^{\mu} q^{\nu}-g^{\mu \nu} q^2) \Pi(q^2)
\,,
\ee
in terms of the correlator of the quark current operator
$J^{\mu}(x)= \bar{q}(x) \gamma^{\mu} q(x) $.
For simplicity we consider one massless quark flavor only.
We examine $D(Q^2)$ in the deep Euclidean region $Q^2\gg\LQ^2$. 

In perturbative QCD,
series expansion of an observable in the
strong coupling constant $\alpha_s$
is expected to be an asymptotic series. 
An IR renormalon is a singularity of the Borel 
transform of a perturbative series on the positive real axis
in the complex Borel plane.
The singularity closest to the origin gives rise to
the leading asymptotic behavior
of the perturbative series.
In the case of $D(Q^2)$, the term of
the asymptotic series becomes minimal at order
$n_*\approx 8\pi/(\beta_0\alpha_s)$
($\beta_0$ is the coefficient of the one-loop beta function of $\alpha_s$)
and grows rapidly beyond that order.
Truncation of the asymptotic series at order $n_*$ induces
a theoretical uncertainty of order $(\LQ/Q)^4$.
On the other hand,
using OPE of the current correlator for large $Q^2$, 
the reduced Adler function can be expressed by the vacuum expectation values (VEVs) 
of gauge and Lorentz invariant local operators as
\be
D(Q^2)=d_1+d_{GG} \frac{\braket{0|G^{\mu \nu a} G_{\mu \nu}^a |0}}{Q^4}+\dots
\,,
\label{ope}
\ee
where $d_1$ and $d_{GG}$ represent the Wilson coefficients
for the operators {\bf 1} and $G^2= G^{\mu \nu a} G_{\mu \nu}^a$,
respectively.  
The (field-dependent) lowest dimension operator $G^2$ has dimension four.
Its VEV is known as the local gluon condensate and
is believed to have a nonzero value of order $\LQ^4$ 
determined by nonperturbative IR dynamics.
This is thought as the origin of the perturbative ambiguity.

\begin{figure}[t]
\begin{center}
\includegraphics[width=0.7\linewidth]{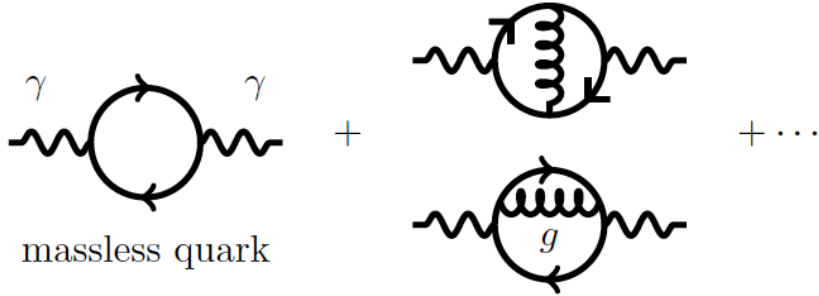}
\vspace*{-2mm}
 \caption{\small{Feynman diagrams for $\Pi(q^2)$.}} 
 \label{diagrams}
 \vspace*{-6mm}
\end{center} 
 \end{figure}

The method of our analysis is as follows.
We first evaluate $D(Q^2)$ in the large-$\beta_0$ approximation
\cite{Beneke:1994qe}.
The leading term stems from the diagrams with
one gluon propagator in Fig.~\ref{diagrams}
(diagrams I).
We consider insertion of a chain of 
one-loop fermion self-energies (with $n_f$ flavors)
to the gluon propagator.
Taking the infinite sum of the chains
and replacing $n_f\to n_f-33/2=-3\beta_0/2$
gives the diagrams I with $\alpha_s(\mu)$
($\mu$ is the renormalization scale in the
$\overline{\rm MS}$ scheme)
replaced by
\begin{align}
&
\alpha_{\beta_0}(\tau)\equiv
\alpha_s^\text{1-loop}(e^{-5/6}\sqrt{\tau})
\nonumber\\&
=\frac{\alpha_s(\mu)}
{1+\frac{\beta_0\alpha_s(\mu)}{4 \pi} \log{(e^{-5/3}\tau/\mu^2)}}
=\frac{4 \pi/\beta_0 }{\log{(e^{-5/3}\tau/\LQ^2)}}
\,,
\end{align} 
where 
$\tau=-k^2$ and $k$ denotes the gluon momentum.
(We set $n_f=1$ in the following.)
Then loop integrals except for the modulus of the (Euclidean) gluon momentum
can be performed, and we obtain the one-dimensional integral 
expression for the reduced Adler function
\cite{Beneke:1994qe,Neubert:1994vb}:
\begin{align}
&D_{\beta_0}^\text{formal}(Q^2)
=\int \!d^4p\, d^4\kappa\ \mathcal{F} (p,\kappa,Q) \frac{\alpha _{\beta_0}(\kappa^2)}{\kappa^2}
\nonumber\\
&=\int ^\infty _{0} \frac{d\tau }{2\pi \tau} 
\alpha _{\beta _0} (\tau )\!
\int\! d^4p\, d^4\kappa\ \mathcal{F} (p,\kappa,Q)\, 2\pi \delta (\tau -\kappa^2)
\nonumber\\
&=
\int_{0}^{\infty} \frac{d \tau}{2 \pi\tau} 
\, w_D\!\biggl(\frac{\tau}{Q^2}\biggr)\, \alpha_{\beta_0}(\tau) \, .
\label{formalser}
\end{align}
Here, $\alpha_s(\mu)\mathcal{F} (p,\kappa,Q)/\kappa^2$ represents the
integrand of the two loop integral expression for the reduced
Adler function, and $p,\kappa$ denote the Euclidean loop momenta
($\tau=\kappa^2=-k^2$).
The above expression is only formal\footnote{
Eq.~(\ref{formalser}) can be made well-defined and precise
by regularization. 
One prescription used extensively is treating the integral as 
the principal-value integral plus a contribution from the Landau pole.
} 
due to existence of the pole at
$\tau=e^{5/3}\LQ^2$ in $\alpha_{\beta_0}(\tau)$ 
and makes sense only in series expansion in $\alpha_s(\mu)$. 
In our formalism, we focus on its UV contributions by introducing an
IR cut-off to the gluon momentum,
\be
D_{\beta_0}(Q^2;\mu_f)\equiv
\int_{\mu_f^2}^{\infty} \frac{d \tau}{2 \pi\tau} 
\, w_D\!\biggl(\frac{\tau}{Q^2}\biggr)\,  \alpha_{\beta_0}(\tau) \,,
\label{redef}
\ee
where the factorization scale is chosen to satisfy $\LQ^2 \ll \mu_f^2 \ll Q^2$.
In this definition the integral path does not include the pole and 
the integral is well-defined.
Although eq.~(\ref{redef}) generally depends on $\mu_f$, we can extract
a $\mu_f$-independent part. 
Such a part is insensitive to IR physics and can be regarded as a
genuine UV contribution.
 
Consider a function $W_D(z)$ which is analytic in the
upper-half complex $z$ plane and satisfies 
\be
2\, {\rm Im} \, W_D(x)=w_D(x) \,\,\,\,\,\, 
(x\in\mathbb{R}~\text{and}~x >0) \,  \label{rel}.
\ee
 Then $D_{\beta_0}$ can be expressed by $W_D$ as
 \be
 D_{\beta_0}(Q^2;\mu_f)={\rm Im} \int_{\mu_f^2}^{\infty} 
\frac{d \tau}{\pi\tau}\, W_D\!\biggl(\frac{\tau}{Q^2}\biggr)\, \alpha_{\beta_0}(\tau)
 \,,
 \ee
and the integral path can be decomposed into the difference 
between $C_a$ and $C_b$ given below. 
\vspace*{3mm}\\
\includegraphics[width=\linewidth]{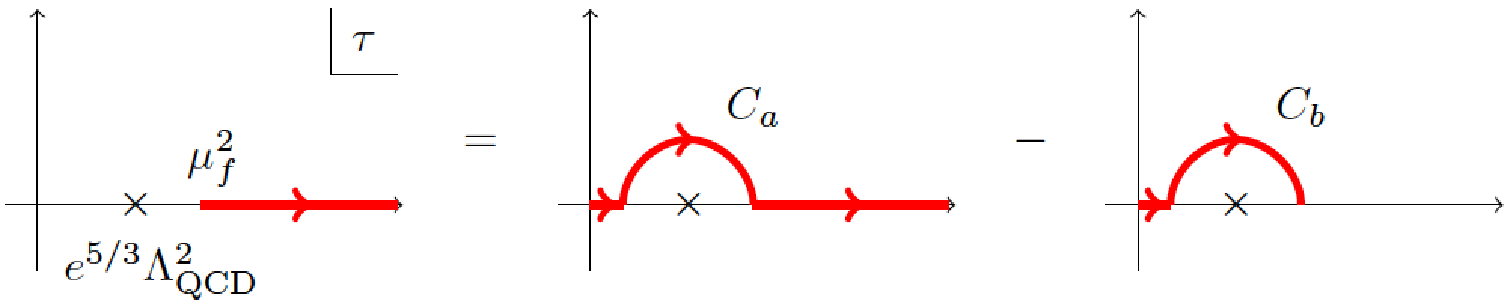}
\vspace*{0mm}\\
The integral along $C_a$ is clearly independent of $\mu_f$. 
The integral along $C_b$ also includes 
$\mu_f$-independent part. 
Since $\mu_f^2\ll Q^2$ it would be justified to
expand $W_D$ about $\tau=0$ along the path $C_b$:
\be
D_{\beta_0}^{(C_b)}(Q^2;\mu_f)=\sum_{n} {\rm Im} \int_{C_b} \frac{d \tau}{\pi\tau} \, c_n \,\biggl(\frac{\tau}{Q^2} \biggr)^n \alpha_{\beta_0}(\tau) \,,
\label{expansion}
\ee
where $W_D (z)=\sum_{n} c_n z^n$. 
For each term,
if $c_n \in \mathbb{R}$, the integral can be written as
\begin{align}
&
{\rm Im} \int_{C_b}\frac{d \tau}{\pi\tau} \,  c_n 
\left(\frac{\tau}{Q^2} \right)^n \alpha_{\beta_0}(\tau)
\nonumber\\ &
=\frac{1}{2\pi i} 
\left(\int_{C_b}-\int_{C^*_b} \right)
\frac{d \tau}{\tau} \,
c_n \left(\frac{\tau}{Q^2} \right)^n \alpha_{\beta_0}(\tau)\,,
\label{eachterm}
\end{align}
since the integrand satisfies the relation $\{ f(z) \}^*=f(z^*)$. 
This reduces to a contour
integral surrounding the pole at 
$\tau=e^{5/3}\Lambda^2_{\rm QCD}$ and the result is its residue:
\be
[\text{eq}.~(\ref{eachterm})]
=-\frac{4 \pi  c_n}{\beta_0} \left(\frac{e^{5/3}\Lambda^2_{\rm QCD}}{Q^2} \right)^n
\,.
\ee 
On the other hand, 
in the case that $c_n$ has a nonzero imaginary part, 
$\mu_f$-dependence generally
remains since the integrand does not satisfy $\{ f(z) \}^*=f(z^*)$. 
In this way $\mu_f$-independent part appears from the integral 
along $C_b$ depending on whether the expansion coefficient is real or complex.

The analytic function $W_D$ which is related to $w_D$ by
eq.~(\ref{rel}) can be constructed systematically. 
If we define 
\be
W_D(z)=\int_{0}^{\infty} \frac{d x}{2 \pi} \frac{w_D(x)}{x-z-i0} 
~~~~~(z\in\mathbb{C})\, ,
\label{WDfromwD}
\ee
it has the desired property. 
Using $w_D$ computed in 
\cite{Neubert:1994vb} we obtain, after appropriate change of conventions,
\begin{align}
 W&_D(z)
 =\frac{N_c C_F}{12\pi} 
 \Bigl[ 3+16z(z+1) H(z)
-14 z^2 \log{(-z)} \nonumber \\ 
&+8 z(z+1) \{ -\log (-z)
{\rm Li}_2(-z)+{\rm Li}_3(z)+{\rm Li}_3(-z)\}\nonumber \\
&+4\{2 z^2+2z+1-4z(z+1)\log{(1+z)}\}{\rm Li}_2(z)  \nonumber \\
& +2(7 z^2-4 z-3)\log{(1-z)}-8\zeta_2z(z+1) \log{(1+z)}
\nonumber \\ 
&+4\{z^2-z(z+1)\log({1+z})\} \log^2{(-z)}
\nonumber \\
&+2(4\zeta_2-7 \zeta_3 )z^2+2(11-7 \zeta_3)z 
\, \Bigr] 
\label{largeW} \,, 
\end{align}
where $N_c=3$ is the number of colors
and $C_F=4/3$ is the Casimir operator of the fundamental representation;
$H(z)=\int_{z}^{1} dx\, x^{-1}\log{(1+x)} \log{(1-x)}$;
${\rm Li}_n(z)=\sum_{k=1}^\infty\frac{z^k}{k^n}$ denotes the
polylogarithm;
$\zeta_k=\zeta(k)$ denotes the Riemann zeta function.\footnote{
$H(z)$ can be expressed by the harmonic polylogarithms.
}

The expansion of $W_D(z)$ in $z$ reads
\begin{align}
 W_D \left(z \right)
 &=N_c C_F \left[\frac{1}{4 \pi}+\frac{8-6 \zeta_3}{3 \pi} z
+\frac{10-12\zeta_3-3 \log{z}+3i\pi}{6 \pi} z^2+\dots \right]  \label{EpdW} \, . 
 \end{align}
 The first two terms have real expansion
coefficients, whereas the third term has a complex coefficient. 
As a result we obtain
\begin{align}
 D_{\beta_0}(Q^2;\mu_f)=
D_{\rm UV}(Q^2)+\mathcal{O}\left({\mu_f^4}/{Q^4}\right) \label{res}
\,,
 \end{align}
with
\begin{align}
&
D_{\rm UV}(Q^2)=D_0(Q^2)+
\frac{8(4\!-\!3 \zeta_3)e^{5/3} N_c C_F}{3 \beta_0}  \frac{\LQ^2}{Q^2}
\,,
\label{defDUV}
\\&
D_0(Q^2)=
\frac{N_c C_F}{\beta_0}+{\rm Im} \int_{C_a} \frac{d \tau}{\pi\tau}
\, W_D\!\biggl(\frac{\tau}{Q^2}\biggr)\,\alpha_{\beta_0}(\tau) \,.
\end{align}
The $\LQ^2/Q^2$ term with the same coefficient has been
obtained in \cite{Ball:1995ni}.
The asymptotic behaviors of $D_0(Q^2)$ can be calculated analytically, which
reads
\be
D_0(Q^2) \to
 \begin{cases}
\frac{N_c C_F}{\beta_0} 
&  \text{as}  \,\,\, Q^2 \to 0  \\
\frac{N_c C_F}{\beta_0} \frac{1}{\log{\left({Q^2}/{\Lambda^2_{\rm QCD}} \right)}} & \text{as} \,\,\, Q^2 \to \infty
 \end{cases} \,.
\label{asympt-Dp}
\ee  
[The behavior as $Q^2\to\infty$ is consistent with the 
renormalization group (RG).]
In the intermediate region both asymptotic forms are interpolated smoothly.
$D_0(Q^2)$ can be easily computed numerically, which is shown
in Fig.~\ref{Fig.Linear}.
$D_0$ has a logarithmic dependence on $\LQ/Q$, which is milder  
compared to power dependences.  
Thus,  
in this way we generate effectively an
expansion in $1/Q^2$, and $D_{\rm UV}$ can be regarded as the
leading terms in this expansion.
(This expansion is not unique due to the
reason discussed below.)

$D_\text{UV}$ is determined only by UV 
contributions and $\mu_f$-independent, i.e.,
insensitive to IR physics.
 Especially $1/Q^2$-term is included and this gives a more dominant contribution 
than $1/Q^4$-term at large $Q^2$. 
In addition, $\Lambda_{\rm QCD}^2$, which cannot 
be expanded in $\alpha_s$, appears together with $1/Q^2$. 
$\mu_f$-dependent terms start from the order $1/Q^4$, which is 
consistent with the fact that $1/Q^4$-term 
has IR contributions in OPE.\footnote{
In the static potential the leading $\mu_f$ dependence (corresponding
to the $\mu_f^4/Q^4$ term in the Adler function) 
cancels against that of the leading nonperturbative matrix element
(non-local gluon condensate) \cite{Brambilla:1999xf,Sumino:2014qpa}.
We expect that a similar cancellation takes place 
also for the Adler function.
To show this explicitly requires
computation of the matrix element in a Wilsonian low-energy effective
field theory with a hard cut-off.
} 
$D_\text{UV}$ is plotted in Fig.~\ref{Fig.Linear} as a function
of $(\LQ/Q)^2$, in which linear dependence
on $(\LQ/Q)^2$ is visible.


In general the correspondence between
perturbative calculation and OPE in
an effective field theory can be
examined using expansion-by-regions
of Feynman diagrams \cite{Beneke:1997zp}.
According to such an analysis 
$D_{\beta_0}$ coincides with $d_1$, since
$D_{\beta_0}$ corresponds to the UV gluon part.
It is clear from the construction that $D_{\beta_0}$ and $D_{\rm UV}$ do
not contain IR renormalons (which stem from the region
$\tau \sim \LQ^2$ \cite{Beneke:1998ui}).
In fact $D_{\rm UV}$ has a well-defined
value (up to a scheme dependence discussed below).

One may suspect that the $\LQ^2/Q^2$ term is an
IR contribution
since it stems from the contour $C_b$ close to the IR pole
at $\tau=e^{5/3}\Lambda _\mathrm{QCD}^2$.
One can verify that this is a UV contribution
using the expansion-by-regions technique.
Combining eqs.~(\ref{formalser}) and (\ref{WDfromwD}), we can
write
\begin{align}
W_D\!\left(\frac{\tau }{Q^2}\right)
=\int\! d^4p\, d^4\kappa\ \frac{\mathcal{F} (p,\kappa,q)}{\kappa^2-\tau -i0}.
\label{RegionNote7}
\end{align}
We separate the momentum regions of
the integral
and investigate them individually.
We use $Q^2$ as a hard-scale parameter
and $\tau \sim \Lambda _\mathrm{QCD}^2$
as a soft-scale parameter.
In the region where all of the quarks and gluon
have hard-scale momenta,
eq.~\eqref{RegionNote7} becomes 
\begin{align}
\left.
W_D\!\left(\frac{\tau }{Q^2}\right)
\right|_\text{all hard}
=\sum _{n=0}^\infty 
\int\! d^4p\, d^4\kappa\ \mathcal{F} (p,\kappa,Q)
\frac{\tau ^n}{(\kappa^2)^{n+1}}.
\label{RegionNote8}
\end{align}
Note that the function $\mathcal{F}$
does not receive any modification
in this region
since the soft-scale parameter $\tau$ is 
contained only in the factor $1/(\kappa^2-\tau )$.
The first and the second ($n=0,1$) terms
of eq.~\eqref{RegionNote8}
exactly reproduce 
the first and the second terms of eq.~(\ref{EpdW}),
respectively.
Therefore we conclude that
the $\Lambda _\mathrm{QCD}^2/Q^2$ term
is a UV contribution.

On the other hand, 
the imaginary part of $c_n$ in eq.~(\ref{expansion})
results in $\mu _f$-dependent terms,
and the $\mu _f$-dependent terms are identified as
IR contributions.
The expansion-by-regions analysis
shows that the imaginary part of $c_n$
indeed stems from the region where
the gluon has a soft-scale momentum.
This is because the only source of imaginary part 
is the integral of $1/(\kappa^2-\tau )$,
namely, once this factor
is expanded as in eq.~\eqref{RegionNote8},
the contribution is explicitly real.

In ref.~\cite{Ball:1995ni}, using the method of massive gluon,
terms which are non-analytic in the gluon mass $\lambda$
are identified as IR contributions,
while terms which are power-like in $\lambda$
as UV contributions.
Written in the form eq.~\eqref{RegionNote7},
it has the same structure as
the massive gluon with a negative mass-squared.
Hence, the source of the imaginary part can be
attributed to the same origin.
For example, a non-analytic term $\ln \lambda ^2$ 
generates an imaginary part when we substitute $\lambda ^2=-\tau$ 
with $\tau >0$.

$W_D$ is not unique because we can add an arbitrary function 
which takes a real value on the positive real axis preserving 
the condition 
(\ref{rel}). 
One can show\footnote{
The variation of 
$D_0$ and that of $c_1$-term, caused by a variation of
$W_D$ ($\delta W_D$), cancel up to 
a residual variation of order $(\LQ/Q)^4$.
Note that $\{ \delta W_D(z) \}^*=\delta W_D(z^*)$.
Furthermore, one can show that the $\mu_f$-dependent
part does not vary.
} that this changes $D_{\rm UV}$ only at order
$(\LQ/Q)^4$ or higher,
although both $D_0$ and coefficient of $(\LQ/Q)^2$ vary.
The reason why the coefficient of $(\LQ/Q)^2$ cannot be determined
uniquely is due to the existence of the $1/\log Q^2$ singularity
in eq.~(\ref{asympt-Dp}) dictated by
RG, which prevents Taylor expansion in $1/Q^2$.
Dependence of $D_{\rm UV}$ on the choice of
$W_D$ may be regarded as a
scheme dependence, tied with the non-existence of
Taylor expansion.

\begin{figure}[t]
 \begin{center}
 \begin{tabular}{cc}
 \hspace*{-4mm}
 \includegraphics[width=0.52\linewidth]{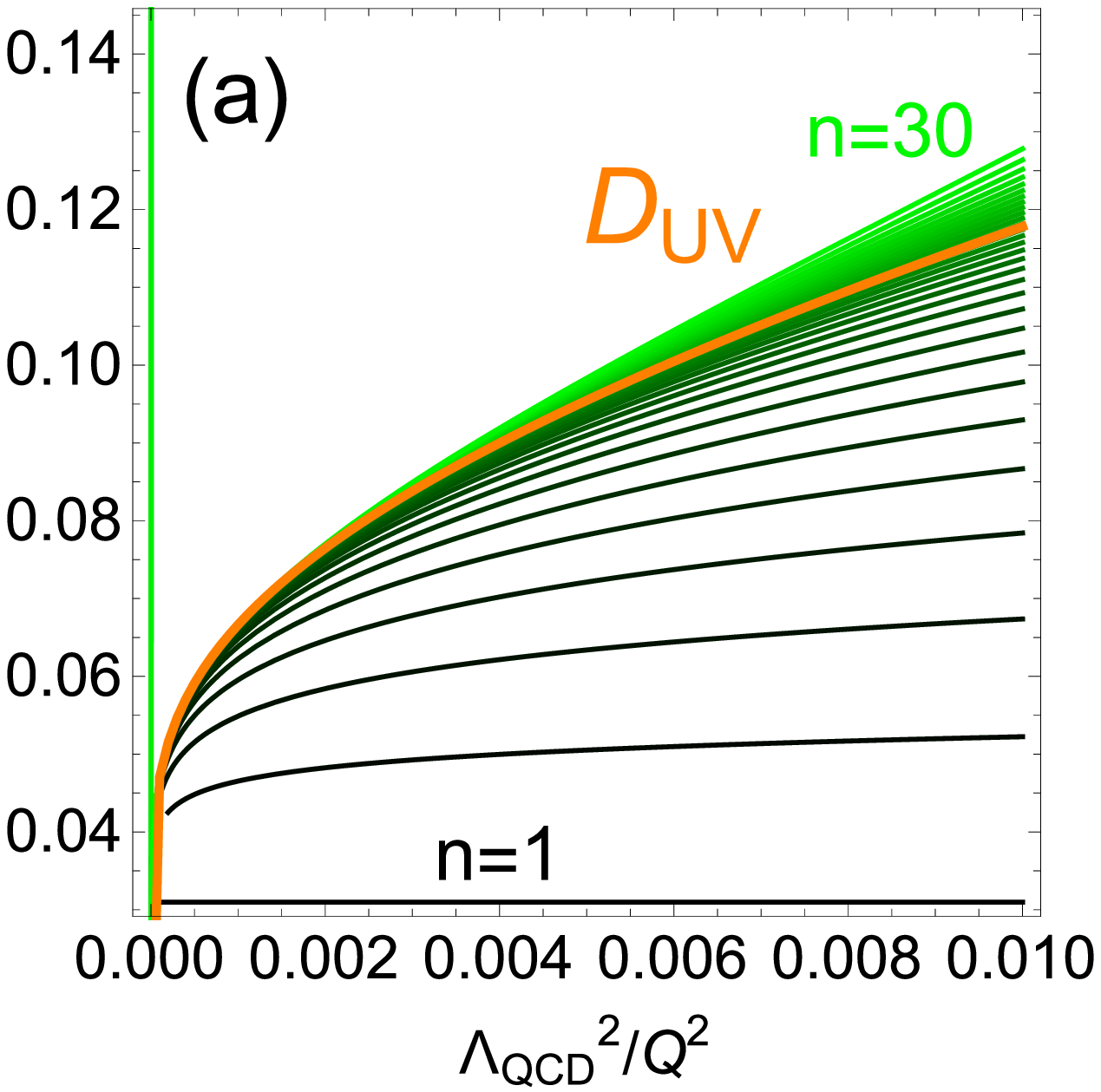}&\hspace*{-4mm}
 \includegraphics[width=0.52\linewidth]{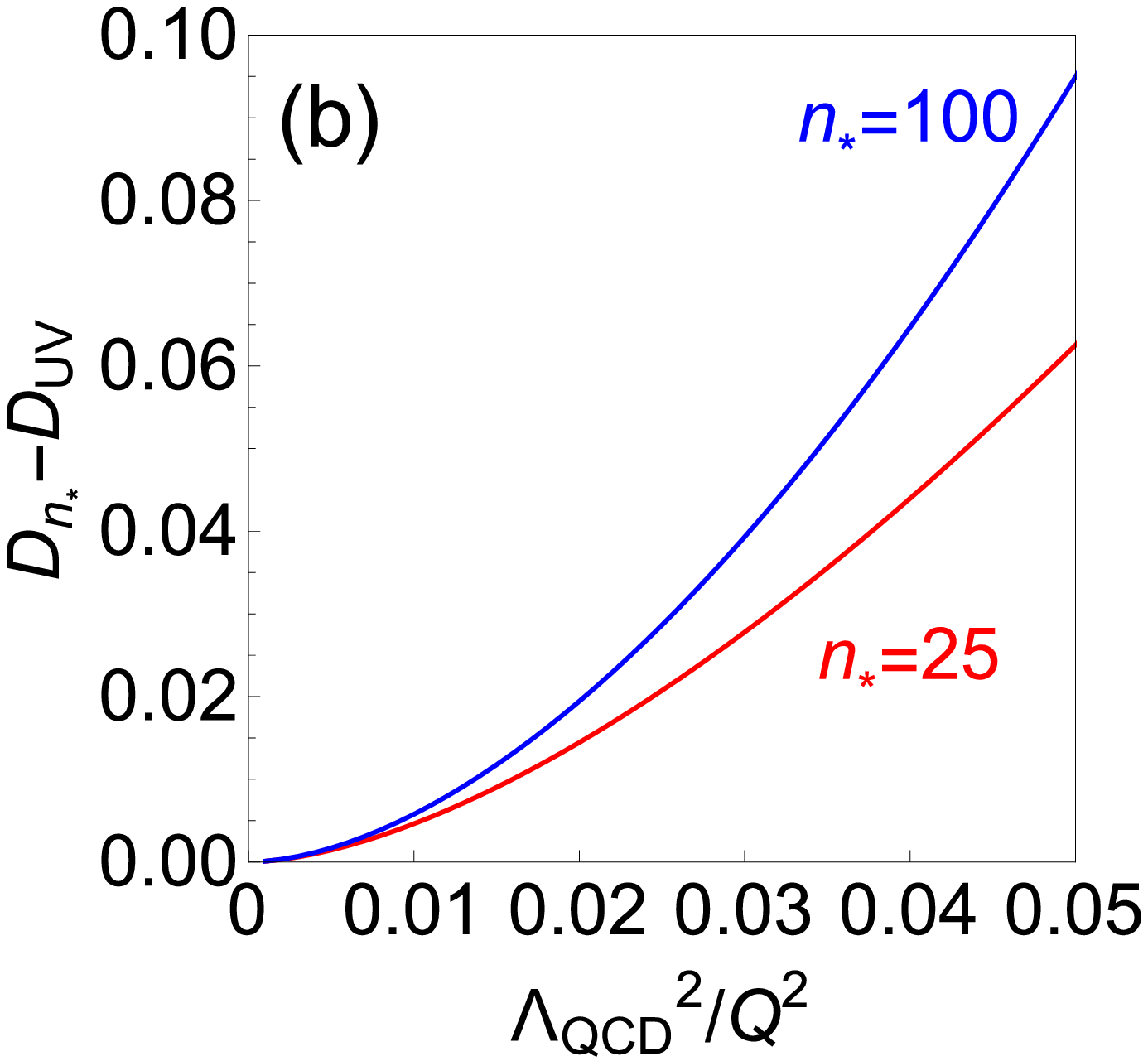}
 \end{tabular}
 \vspace*{-2mm}
 \caption{\small 
(a)
Sum of series expansion of $D_{\beta_0}^\text{formal}$
up to ${\cal O}(\alpha_s^{n})$ and $D_\text{UV}$.
The input is $\alpha_s(\mu)=0.097$ (corresponding to $n_*=25$).
(b) 
$D_{n_*}-D_\text{UV}$ vs.\ $\LQ^2/Q^2$, for $n_*=25$ and 100.
} 
 \label{difference}
 \vspace*{-5mm}
\end{center} 
 \end{figure}

We can also extract the same $D_{\rm UV}$
by truncating the formal series expansion (\ref{formalser})
at order $n_*\approx 8\pi/(\beta_0\alpha_s(\mu))$
($D_{n_*}$)
and examining the limit $\alpha_s(\mu)\to 0$,
following the analysis method in \cite{Sumino:2003yp}.
As $\alpha_s(\mu)\to 0$ ($n_*\to \infty$)
the truncated series approaches $D_{\rm UV}$
up to a slowly diverging ${\cal O}(\LQ^4/Q^4)$ part,
i.e., $D_{n_*}-D_\text{UV}\sim \log n_* \times {\cal O}(\LQ^4/Q^4)$
for $n_*\gg 1$.
In Fig.~\ref{difference}(a) we compare
$D_\text{UV}$ and the sum of the truncated series
up to ${\cal O}(\alpha_s^{n})$ 
in the case $\alpha_s(\mu)=0.097$ (corresponding to $n_*=25$).
We show in Fig.~\ref{difference}(b) 
the difference 
$D_{n_*}-D_\text{UV}$ for the choices of $\alpha_s(\mu)$
corresponding to $n_*=25$ and 100.
The difference for each $n_*$ indeed behaves as ${\cal O}(\LQ^4/Q^4)$, 
namely it reduces at larger $Q^2$. 

\begin{figure}[h!]
 \begin{center}
 \includegraphics[width=1\linewidth]{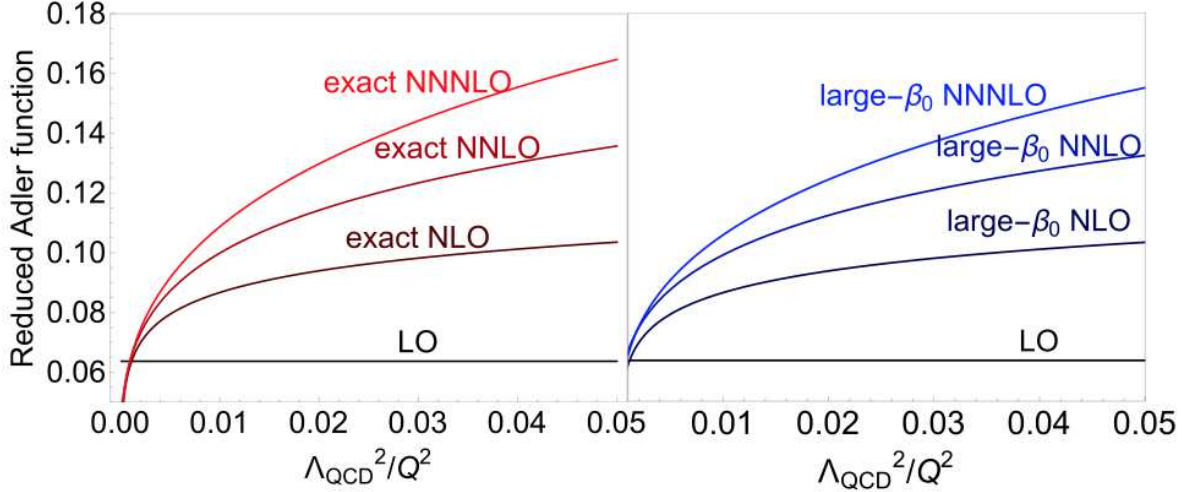}
 \vspace*{-4mm}
 \caption{\small{Perturbative series of $D(Q^2)$: 
exact result for the non-singlet part (left) and large-$\beta_0$
approximation (right).
N$^k$LO line represents the sum of the series up to ${\cal O}(\alpha_s^{k+1})$. 
The input is taken as $\alpha_s(\mu)=0.2$.}} 
 \label{fixed_order}
 \vspace*{-5mm}
\end{center} 
 \end{figure}

So far we have analyzed in the large-$\beta_0$ approximation. 
The exact perturbative series is known up to 
$\mathcal{O}(\alpha_s^4)$  \cite{Baikov:2008jh}. 
We compare both results up to this order in Fig.~\ref{fixed_order}
and see qualitatively a good agreement.
Hence,
our analysis in the large-$\beta_0$ approximation looks consistent.
Namely, it suggests that at large ($\lesssim n_*$) order 
of perturbative expansion linear dependence on $(\LQ/Q)^2$
would appear;
see Fig.~\ref{difference}(a).

Thus, we have extracted UV contributions to the Adler function,
free from IR renormalons,
and presented an understanding of the
$\LQ^2/Q^2$ term included in it. 
Namely, the leading
Wilson coefficient $d_1$  can be identified with $D_{\rm UV}$
up to order $\mu_f^4/Q^4$, where
$D_{\rm UV}$ is given as a sum of a logarithmic term $D_0$
and a $\LQ^2/Q^2$ term.
We showed that the $\LQ^2/Q^2$ term is indeed included in large order perturbative series.
Note that this power behavior is different from a perturbative uncertainty 
induced by the UV renormalon located on the negative real axis in
the Borel plane. 
The contribution from UV renormalon is renormalization scale dependent 
and becomes less important as we raise the order of truncation ($n_*$) properly.  
Note also that the separation into the $D_0$ and $\LQ^2/Q^2$ terms is not unique.
It is difficult to eliminate
this dependence 
by expansion in $1/Q^2$ due to $1/\log Q^2$ singularity
in the leading term.
This scheme dependence is {\it not} a perturbative uncertainty
and eventually cancels in $D(Q^2)$.
From comparisons in Figs.~\ref{Fig.Linear}
and \ref{difference}  we find the
scheme choice given by eq.~(\ref{WDfromwD})
a reasonable one.
In this way the $\LQ^2/Q^2$ term is intrinsic to the
perturbative prediction of the Adler function.


In the case of the static potential a method of
systematic improvement (beyond large-$\beta_0$ approximation)
was devised and applied,
which resulted in
a better agreement with lattice results 
\cite{Sumino:2003yp,Sumino:2014qpa}.
Unfortunately the same method does not work for the
Adler function.
Nevertheless in principle any improvement in
the UV region justifiable in perturbative QCD
should be valid since our method depends only on this part.


Finally we remark
that the formulation presented in this Letter 
would be applicable to more
general observables, at least to those which depend only on one scale
and in the large-$\beta_0$ approximation.

The works of G.M.\ and Y.S.\ are supported in part by JSPS KAKENHI 
Grant Number14J10887 and by Grant-in-Aid for
scientific research (No.~26400238) from
MEXT, Japan, respectively.





\end{document}